\documentclass[12pt]{iopart}
\usepackage{graphicx}
\usepackage{amssymb}

\newcommand{\Qvec}{{\mathbf{Q}}}
\newcommand{\nvec}{{\mathbf{n}}}
\newcommand{\lvec}{{\mathbf{l}}}
\newcommand{\mvec}{{\mathbf{m}}}
\begin{document}

\title[The zigzag disclination]{Nematic liquid crystals on sinusoidal channels: the zigzag instability}

\author{Nuno M. Silvestre$^{1,2,*}$, Jose M. Romero-Enrique$^3$, and Margarida M. Telo da Gama$^{1,2}$}

\address{$^1$ Departamento de F{\'\i}sica da Faculdade de Ci\^encias, Universidade de Lisboa, Campo Grande, P-1649-003 Lisboa, Portugal.}
\address{$^2$ Centro de F{\'\i}sica Te\'orica e Computacional, Universidade de Lisboa, Campo Grande, P-1649-003 Lisboa, Portugal.}
\address{$^3$ Departamento de F\'{\i}sica At\'omica, Molecular y
Nuclear, Area de F\'{\i}sica Te\'orica, Universidad de Sevilla,
Apartado de Correos 1065, 41080 Sevilla, Spain}
\ead{nmsilvestre@ciencias.ulisboa.pt}
\vspace{10pt}
\begin{indented}
\item[Submitted: ]July 2016
\end{indented}

\begin{abstract}

Substrates which are chemically or topographically patterned induce a variety of liquid crystal textures. The response of the liquid crystal to competing surface orientations, typical of patterned substrates, is determined by the anisotropy of the elastic constants and the interplay of the relevant lengths scales, such as the correlation length and the surface geometrical parameters. Transitions between different textures, usually with different symmetries, may occur under a wide range of conditions.
We use the Landau-de Gennes free energy to investigate the texture of nematics in sinusoidal channels with parallel anchoring bounded by nematic-air interfaces that favour perpendicular (hometropic) anchoring. In micron size channels 5CB was observed to exhibit a non-trivial texture characterized by a disclination line, within the channel, which is broken into a zigzag pattern. 
Our calculations reveal that when the elastic anisotropy of the nematic does not favour twist distortions the defect is a straight disclination line that runs along the channel, which breaks into a zigzag pattern with a characteristic period, when the twist elastic constant becomes sufficiently small when compared to the splay and bend constants. The transition occurs through a twist instability that drives the defect line to rotate from its original position. The interplay between the energetically favourable twist distortions that induce the defect rotation and the liquid crystal anchoring at the surfaces leads to the zigzag pattern. 
We investigate in detail the dependence of the periodicity of the zigzag pattern on the geometrical parameters of the sinusoidal channels, which in line with the experimental results is found to be non-linear.  

\end{abstract}

%
%
%
%
%

\section{Introduction}
\label{sec:intro}

Nematic liquid crystals are states of matter with long-ranged orientational order and positional
disorder \cite{Gennes.1993}. However, the presence of boundaries or external fields can frustrate this order,
leading to elastic distortions and the formation of topological defects.
In particular, the alignment of nematic liquid crystals induced by substrates is crucial for practical applications, as
display devices. In this respect, the topography of the substrate plays an essential role to determine the nematic alignment.
For example, periodic grooved surfaces have been studied extensively in the literature due to the possibility of obtaining
coexisting stable nematic textures with different optical properties, important for the development of zenithal
bistable devices \cite{Brown_et_al_2000,Uche_et_al_2005,Uche_et_al_2006,Davidson_et_al_2010,Evans_et_al_2010,Dammone_et_al_2010,Raisch.2014}. Their power consumption is much lower than that of standard liquid crystal displays since they require power only
to switch between the different textures. In most cases, the nematic director is assumed to have translational symmetry
along the groove axis. This condition is taken as granted when the interfaces align the nematic molecules in directions perpendicular to the grooves axis. However, recent results indicate that the translational
symmetry may be broken. For example, liquid crystals on sinusoidal microwrinkle grooves may develop a zigzag defect line
\cite{Ohzono.2012}. This new periodic structure along the groove axis is associated to twist distortions of the nematic
texture which are absent on the zenithal arrangement, and their origin can be traced to the fact that twist is
energetically more favourable than bend or splay distortions \cite{Ohzono.2012,Buscaglia.2010}. A similar phenomenon has
been observed in achiral nematic liquid crystals confined in cylindrical pores with homeotropic anchoring
\cite{Jeong_et_al_2015}. When the cylindrical pore has random planar anchoring, double-twist chiral structures are also
observed \cite{Davidson_et_al_2015,Nayani_et_al_2015}. In this case, the driving force is the (surface) saddle-splay
contribution to the free energy, which favours director alignment along the direction of highest surface curvature.
Finally, we note that the spontaneous chiral symmetry has also been predicted and observed in bipolar
\cite{Volovik_Lavrentovich_1983,Williams_1986,Vanzo_2012} and toroidal \cite{Pairam_et_al_2013} droplets.

In this paper we will perform a numerical study within the Landau-de Gennes framework of the nematic texture of a filled
groove or channel of a substrate with sinusoidal relief, which mimics the experimental setup from Ref. \cite{Ohzono.2012}. Nematics
in contact with sinusoidal grooves have been extensively studied from a theoretical and numerical point of view
\cite{Raisch.2014,Berreman.1972,Fukuda_Yoneya_Yokoyama_2007,Barbero_etal_2008,Patricio_Telo_Dietrich_2002,
Patricio.2011,Li_Zhang_2015}, but as far as we know there is no detailed numerical studies which reproduce
the experimental observations from Ref. \cite{Ohzono.2012}. This is the main goal of our paper: to obtain
theoretically the conditions under which the zigzag disclination line is observed, as well as to characterize
the fine structure of the disclination line. This study will be a starting point to considering
more complex situations. This paper is organized as follows. We describe the model and numerical methodology in Section
\ref{sec:model}. The discussion of our results is presented in Section \ref{sec:results}. We end up the paper with the
conclusions in Section \ref{sec:conclusions}.

\section{Model}
\label{sec:model}


The nematic liquid crystal is modelled by the Landau-de Gennes (LdG) mean-field theory \cite{Gennes.1993}, where the local orientational order is described by a traceless, symmetric, tensor order parameter $\Qvec$. 
In the general case, $Q_{ij}=S(3n_in_j-\delta_{ij})/2 + B(l_il_j-m_im_j)/2$, where the first term measures the uniaxial order while the second becomes non-zero when biaxiality is present. $\nvec$ is the 
nematic director field and $S$ is the scalar order parameter (degree of orientational order). $B$ is the biaxial order parameter. $\lvec$ and $\mvec$ are orthonormal vectors; together with $\nvec$ they make a local basis.  
The free energy is written in terms of invariants of $\Qvec$ and its derivatives,
 $F=\int_V{d^3x\left(f_b\left(\Qvec\right)+f_e\left(\partial\Qvec\right)\right)}$, where the bulk and the elastic free energy densities are, respectively,
 \begin{eqnarray}
 f_b&=&a_o\left(T-T^*\right)\Tr \Qvec^2-b\Tr\Qvec^3+c\left(\Tr\Qvec^2\right)^2\\
 f_e&=&\frac{L_1}{2}\partial_\gamma Q_{\alpha\beta}\partial_\gamma Q_{\beta\alpha} 
 + \frac{L_2}{2}\partial_\gamma Q_{\alpha\gamma}\partial_\delta Q_{\delta\alpha}.
 \end{eqnarray}
The bulk term $f_b$ controls the nematic-isotropic phase transition, and sets the preferred value for the scalar order parameter; $S_{bulk}=0$ in the isotropic phase, and $S_{bulk}=\left(b/8c\right)\left(1 + \sqrt{1-8\tau/9}\right)$ in the nematic. $\tau=24a_o(T-T^*)c/b^2$ is a reduced temperature. The elastic term $f_e$ penalizes distortions of the LC 
director (orientational) field. 
We use two elastic constants $L_1$ and $L_2$ that are related to the Frank-Oseen elastic constants (for splay, bend, and twist distortions) by $K_{splay}=K_{bend}=9S_{bulk}^2L_1\left(2+L_2/L_1\right)/4$ and $K_{twist}=9S_{bulk}^2L_1/2$.  

For the liquid crystal 5CB in the nematic phase the LdG parameters are \cite{Kralj.1991} $a_o=0.044$ MJ K$^{-1}$ m$^{-3}$ , $b=0.816$ MJ m$^{-3}$, $c=0.45$ MJ m$^{-3}$, $L_1=6$ pJ m$^{-1}$, $L_2=12$ pJ m$^{-1}$, and $T^*=307$ K. 
We have set the temperature at $T=307.2$ K, just below the isotropic-nematic transition temperature, $T_{\mathrm{IN}}(\mathrm{5CB})=308.5$K.


\begin{figure}[t]
\center
\includegraphics[width=0.9\columnwidth]{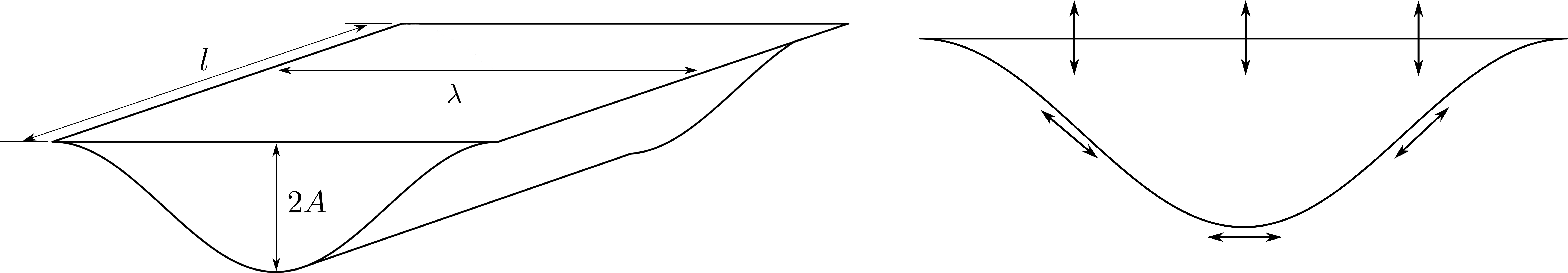}
\caption{Illustration of the geometrical parameters (left) and boundary conditions (right). The sinusoidal channels have wavelength $\lambda$ and amplitude (peak-to-peak) $2A$. $l$ is the length of the simulation box. The preferred alignment 
of the nematic director $\nvec$ is parallel to the sinusoidal surface, and perpendicular at the interface (top boundary).}
\label{fig:geometry}
\end{figure}

We consider a nematic film that fills the sinusoidal channel (see Fig.\ref{fig:geometry}), of wavelength $\lambda$ and peak-to-peak amplitude $2A$. The channel imposes tangential (parallel) anchoring along the sinusoidal channel. A nematic-air interface promotes perpendicular anchoring and is pinned at the crests, capping the channel. We assume that the interface is rigid, and that on both the substrate and the interface the liquid crystal alignment is fixed (Dirichlet boundary conditions). We use free (Neumann) boundary  conditions at the ends along the channel.

We minimize the Landau-de Gennes free energy numerically, using finite elements (FEM) with a quasi-Newton method and adaptive meshes. Details of the numerical method may be found in \cite{Tasinkevych.2012}. The numerics have an accuracy  better than $1\%$. We note that the phenomena discussed here occurs at the micrometer length scales, over two orders of magnitude larger than the correlation length, which sets the size (width) of the topological defect (line). As a consequence the numerical calculations are extremely lengthy, and for the largest systems full minimization took over 3 months of CPU time using AMD Opteron $^{TM}$ 6376 (2.5GHz) processors.

\section{Results and discussion}
\label{sec:results}

\begin{figure}[t]
\center
\includegraphics[width=0.9\columnwidth]{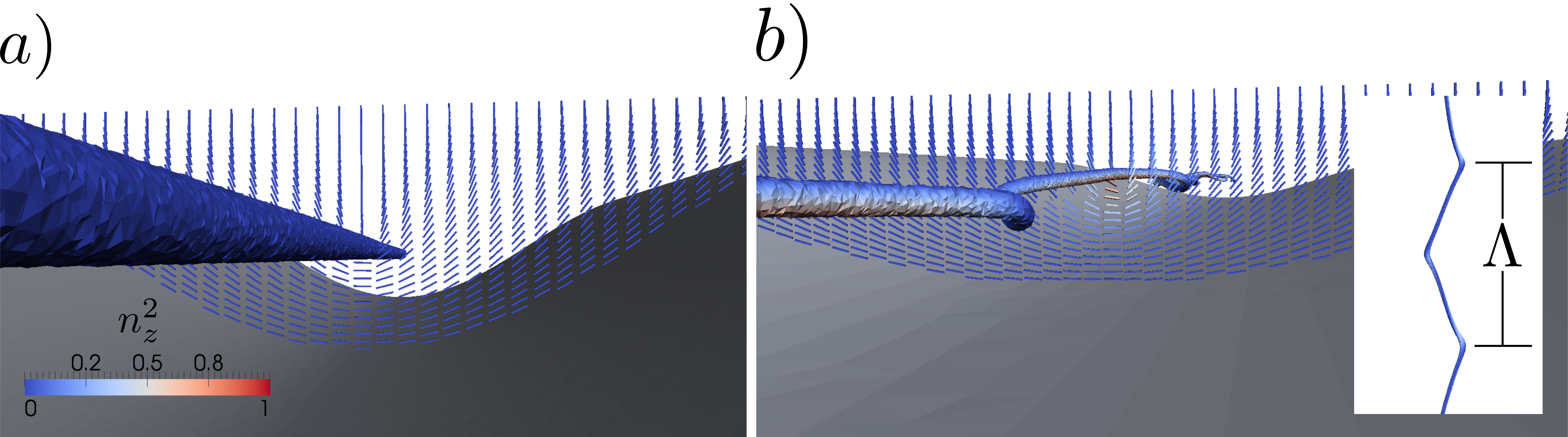}
\caption{ Half-interger topological defect lines in sinusoidal channels. a) Straight line defect and b) zigzag defect along the sinusoidal channel. The inset shows the top view of the zigzag defect. $\Lambda$ is the period of the zigzag. The bar represents the nematic director $\nvec$ half-way in the channel. The color code denotes $n_z^2$; red ($n_z^2=1$) if the nematic points along the channel, and blue ($n_z^2=0$) otherwise.}
\label{fig:defects}
\end{figure}

\begin{figure}[t]
\center
\includegraphics[width=0.6\columnwidth]{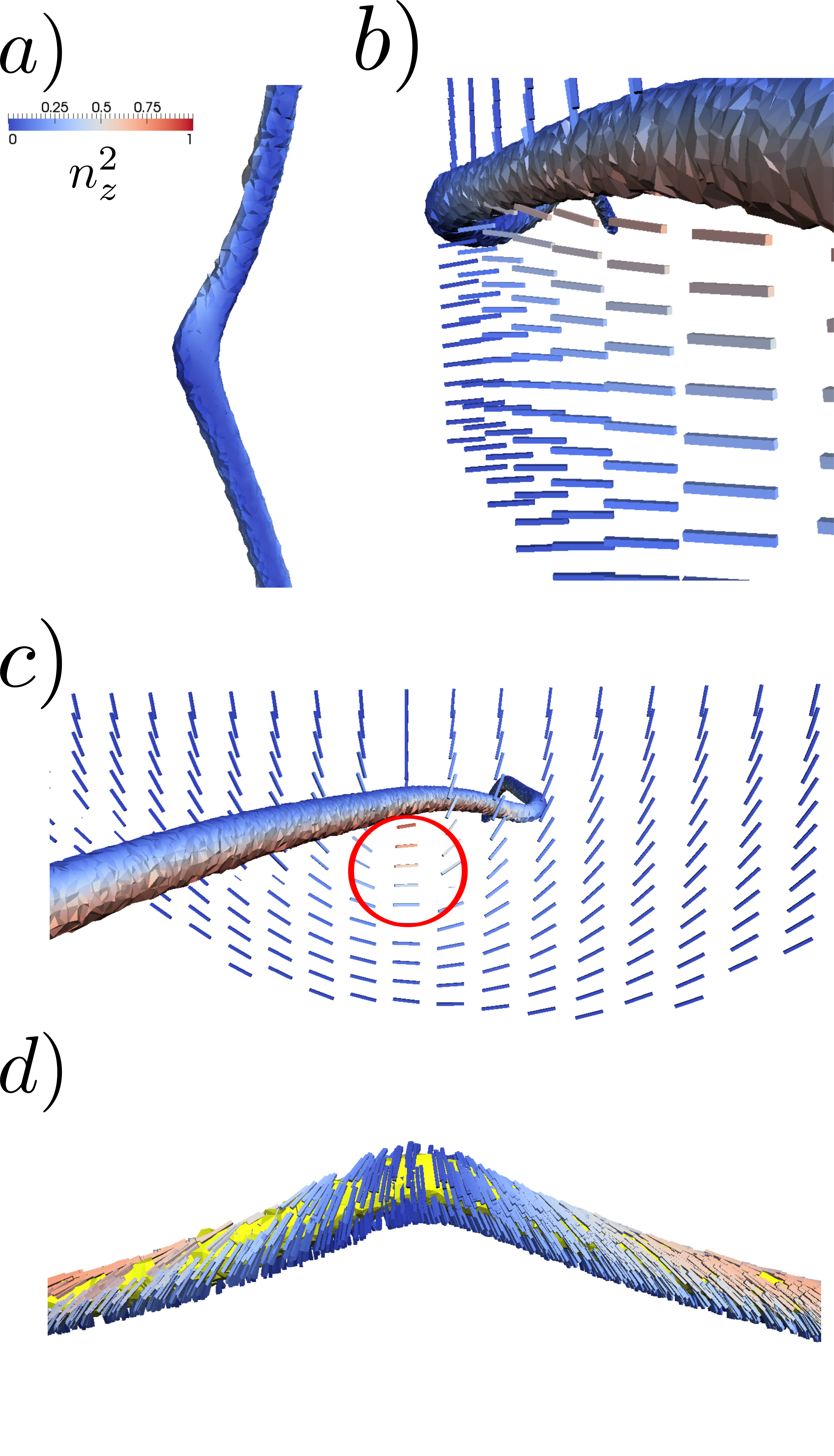}
\caption{ Configuration of the zigzag defect. a) Top view of the defect kink. b) 
Configuration of the nematic along the defect. Just below the defect the director $\nvec$ exhibits a twist distortion. c) Configuration of the nematic on a plane perpendicular to the sinusoidal channel. Below the defect the nematic twists from the orientation imposed by the substrate. The red circle indicates the region where the twist distortion is evident. d) Bottom view of the defect (in yellow) with the nematic configuration around it. At the kink the nematic splays to change its orientation. The nematic director $\nvec$ is represented by the bars. The color code denotes $n_z^2$; red ($n_z^2=1$) if the nematic points along the channel, and blue ($n_z^2=0$) otherwise.}
\label{fig:zigzag}
\end{figure}

For their application in current technology (e.g., electronic shelf labels), nematic cells with sinusoidal substrates have attracted much attention. They constitute a class of simple zenithally bistable devices (ZBD) whose development was driven by the significant interest in controlling the liquid crystal texture through surface patterning. 
The idea that geometry can affect the liquid crystal alignment comes from the seminal work of Berreman \cite{Berreman.1972}, where he showed that a sinusoidal substrate (at low roughness) could induce a preferred orientation of the nematic director. 
This result has been extended to the large roughness case by subsequent works \cite{Fukuda_Yoneya_Yokoyama_2007,Barbero_etal_2008}.   

Patterned substrates impose elastic distortions that can be accommodated in different ways, depending on the elastic constants of the liquid crystal. They also increase the number and complexity of the stable configurations or nematic textures. For example, it was demonstrated that nematic cells with (at least) one sinusoidal substrate can accommodate three equilibrium states that differ by the existence and type of the defects in the nematic matrix \cite{Raisch.2014}. Their influence is also significant in wetting phenomena. We have studied how sinusoidal patterns affect wetting of a substrate by a nematic \cite{Patricio.2011} and shown that the wetting state (with the nematic-isotropic interface far from the substrate) is suppressed at high roughness being replaced by a filled state (with the interface pinned at the top of the channels) to avoid the nucleation of topological defects.

In all of these cases the system is, or is assumed to be, translational invariant. This means that when a defect appears it is a line that runs along the channel. An example is given in Fig.\ref{fig:defects}a. The color code indicates that there is no component of the director field along the channel $n_z=0$. It is a $+1/2$ straight disclination line that appears due to orientational frustration due to the competition between homeotropic alignment at the interface (top boundary) and tangential alignment at the substrate. However, under conditions discussed below, the line defect may not be straight  breaking the translational invariance of the system.

Recent experiments revealed that under some conditions the defect line is no longer straight but zigzags along the sinusoidal channel \cite{Ohzono.2012}. In the experiment, a nematic 5CB film is spread on sinusoidal channels of wavelength $0.6 < \lambda < 20 \,\mu$m and height to wavelength ratio that is nearly constant, $A/\lambda=0.13$. For periods shorter than $\lambda\sim 1.5\mu$m the nematic film exhibits a topological defect, which is a straight disclination line (Fig.\ref{fig:defects}a), while for $\lambda > 1.5\mu$m a stable periodic structure appears. Optical measurements indicate the presence of a zigzag defect line (Fig.\ref{fig:defects}b). In the limit of very shallow channels the elasticity dominates the surface alignment and the nematic texture is nearly uniform, free from topological defects. 

The zigzag line appears as the straight line defect becomes unstable. The instability is driven by the elastic anisotropy of the liquid crystal. Around the $+1/2$  topological defect the nematic director exhibits splay and bend distortions. As discussed in \cite{Buscaglia.2010}, when the twist elastic constant is smaller than the splay and bend elastic constants the nematic releases energy by twisting around the line defect. Using a simplified model, Buscaglia {\it et al} \cite{Buscaglia.2010} have shown that when $K_{twist}<\left(K_{splay}+K_{bend}\right)/2$ the nematic director close to the defect, in the region where the twist distortion occurs, aligns along the defect line. 

In Fig.\ref{fig:defects}b we depict the configuration of a zigzag defect obtained through the numerical minimization described in Sec.\ref{sec:model}. Clearly, the defect is no longer a straight line. The inset shows the top view. The straight defect line was broken into pairs of segments oriented at symmetric angles joined at a kink, where the orientation of the segments changes. This structure was indeed suggested in \cite{Ohzono.2012} as the result of the twist instability discussed in \cite{Buscaglia.2010}, where each segment corresponds to a twist distortion with a given handedness that alternates along the line.

Figure \ref{fig:zigzag} shows in more detail the structure of the zigzag defect, particularly in the region close to the kink. The bars represent the nematic director and illustrate the orientation of the liquid crystal. The color code denotes the component of the director along the channel $n_z$; red ($n_z^2=1$) if the director is along the channel, blue ($n_z^2=0$) if the director is perpendicular. In the region between the defect and the nematic-air interface the director takes the homeotropic alignment imposed by the top boundary, while close to the sinusoidal substrate the nematic takes the tangential anchoring imposed by the channel surface. Along a large closed path encircling the defect, the director configuration is similar to that of the straight line defect (shown in Fig.\ref{fig:defects}a), consistent with a winding number (on the cross-section) $+1/2$. However, in the region just below the defect (see Fig.\ref{fig:zigzag}c) the nematic exhibits a twist distortion. Fig.\ref{fig:zigzag}b shows the nematic director on a vertical plane tangent to one of the defect segments . The director twists from the bottom to the top of the channel, reaching the defect with nearly the same orientation as the segment itself. This confirms the results of 
Buscaglia {\it et al} who found that the director aligns along the defect \cite{Buscaglia.2010}. This also indicates that the twist distortion induces the defect to deviate from its straight configuration along the channel and confirms that the mechanism for the zigzag is a twist instability.  
In other words, it is favourable to rotate the defect line to accommodate the energetically lower twist distortions between the defect and the substrate. 
However, the distance between the defect and the sinusoidal substrate decreases as we move along one segment and the surface anchoring eventually wins preventing the nematic from twisting. 
As a result the line defect changes direction, through a kink, towards the center of the channel where it can accommodate twist distortions. The kink, where two regions with twist distortions of different handedness merge, is characterized by a splay distortion as shown in Fig.\ref{fig:zigzag}d.

\begin{figure}[t]
\center
\includegraphics[width=0.9\columnwidth]{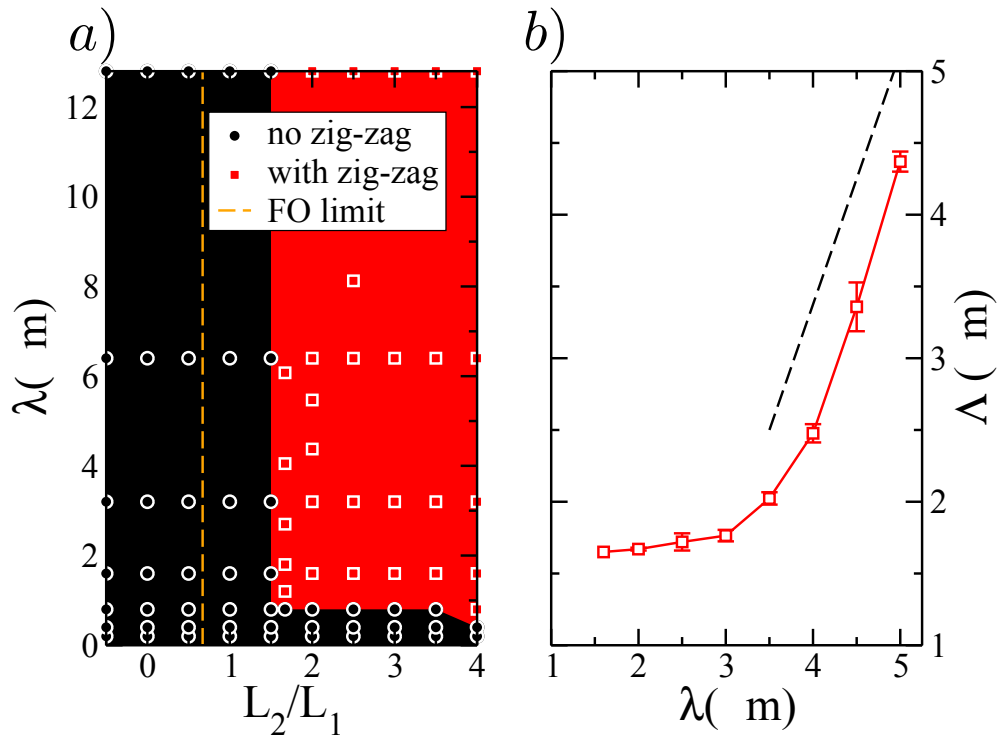}
\caption{ a) Configuration diagram represented as the wavelength of the sinusoidal channels $\lambda$ {\it versus} the elastic anisotropy $L_2/L_1$. In red is the region where the defect line adopts a zigzag configuration; in black is the region where the defect line is straight. The dashed line indicates the limit of stability obtained from the Frank-Oseen theory. b) Period of the zigzag defect line $\Lambda$ as a function of the wavelength of the sinusoidal channel $\lambda$. The dashed line corresponds to $\Lambda\sim1.75\lambda$. }
\label{fig:diagram}
\end{figure}

The twist distortion appears when the twist elastic constant is lower than the splay and bend elastic constants. In Ref.\cite{Ohzono.2012} the authors used the Frank-Oseen theory to estimate the limit of stability of the straight line defect, in the strong anchoring regime. They have found that the instability occurs when 
\begin{equation}
\frac{K_{twist}{\cal A}/d^2}{\frac{\pi}{4}K\log{\left(R/r_c\right)}}<\frac{1.1\left(1-K_{twist}/K\right)}{0.1+K_{twist}/K},
\label{eq:threshold}
\end{equation}
where ${\cal A}$ is the area under the defect where the twist distortion occurs, and $d$ is the vertical diameter of that region. $K=K_{splay}=K_{bend}$ is the value of the splay and bend elastic constants, assumed to be equal. $r_c$ is the defect core radius, and $R$ is the radius of a cylinder surrounding the defect and delimiting the region where strong elastic distortions occur. Assuming ${\cal A}\simeq\pi\left(d/2\right)^2$ and $R\simeq 10 r_c$ the Frank-Oseen elastic theory indicates that the straight line defect becomes unstable when $K_{twist}<0.75K$. From the relation between the Frank-Oseen and the Landau-de Gennes elastic constants (see Sec.\ref{sec:model}) this implies that the zigzag line will appear at $L_2>0.67 L_1$.

In order to investigate the stability of the zigzag defect we performed numerical simulations in sinusoidal channels with a free nematic interface, with boundary conditions described in Sec.\ref{sec:model}. The wavelength of the channels $\lambda$ takes values between $0.2\mu$m and $12.8\mu$m. The length of the simulation box was not smaller than $l=2\lambda/3$, and the height of the channels $2A=0.2\lambda$. We varied the elastic anisotropy $L_2/L_1$ over a wide range; we took $L_1=6$pJ m$^{-1}$ and thus the system with $L_2=2L_1$ corresponds to 5CB.
The stability of the nematic textures was investigated by starting from different initial conditions corresponding to the straight or the zigzag defects, and comparing the free energies of the final configurations.

Figure \ref{fig:diagram}a summarizes the configuration diagram. In red is the region where we found the zigzag to be stable, and in black is the region where the defect is a straight line. For comparison we also plot the Frank-Oseen stability limit discussed above. Our numerics indicate that the zigzag defect appears when $L_2/L_1\gtrsim1.5$, which is more than twice the value predicted by the Frank-Oseen theory, but still agrees with experiment \cite{Ohzono.2012}. The discrepancy between our results using the Landau-de Gennes theory and the Frank-Oseen theory can, on one hand, arise from the (uncontrolled) approximations made in the latter, such as the size of the cylindrical region enclosing the defect $R$ and the area where the twist distortions occur ${\cal A}$. On the other hand, the Frank-Oseen threshold given by Eq.\ref{eq:threshold} neglects the geometric details of the substrate. For example, we have seen in our numerics that increasing the height of the channels at fixed wavelength stabilizes the straight line defect. 

The configuration diagram also shows that, above the elastic anisotropy stability threshold, the zigzag appears only in channels with wavelength $\lambda\gtrsim0.8\mu$m, which is below the experimental threshold $1.5\mu$m. This difference may arise from irregularities on the surface of the channel, as the zigzag only appears if the area between the defect and the substrate can accommodate smooth twist distortions.

The periodicity of the zigzag is the result of the energetic balance between the cost of maintaining the twist distortion as the defect approaches the substrate and that of forming a kink. At large values of $\lambda$ ($>5\mu$m) the periodicity is found to depend linearly on the wavelength $\Lambda\propto \lambda$ \cite{Ohzono.2012}. However, this dependence becomes non-linear at small values of $\lambda$, an indication that there is more than one relevant length scale in the problem.
 
In order to determine the dependence of $\Lambda$ on $\lambda$ we adopted the following procedure.
We set the wavelength $\lambda$. We fix the elastic anisotropy at $L_2/L_1=2$ and the height to wavelength ratio $2A/\lambda=0.2$. We then minimized the free energy for different box lengths $l$. The free energy for each box has two main contributions $F=F_o+F_d$. The first is a contribution that arises from increasing, or decreasing, the channel length by an amount $\Delta l$, and grows linearly with $\Delta l$. The second contribution is associated to the actual zigzag distortion and increases as the box length $l$ becomes incommensurate with the zigzag periodicity. 
By considering several simulations boxes, with different lengths, and for configurations of the same type (i.e., with the same number of kinks) we can find the size $l_{min}$ that minimizes the free energy for a given wavelength $\lambda$ and, as a result, obtain the periodicity of the defect. 
 
Our results for $\Lambda(\lambda)$ are summarized in Fig.\ref{fig:diagram}b. At very small values of the wavelength $\lambda\sim 1.5\mu$m the periodicity of the zigzag approaches a constant value $\Lambda\sim1.65\mu$m. As the wavelength is increased the lateral distance available for the defect to zigzag also increases and the periodicity grows slowly until $\lambda\sim 4\mu$m, after which a linear behaviour $\Lambda\propto\lambda$ is obtained. For comparison, in Fig.\ref{fig:diagram}b the dashed line corresponds to $\Lambda\sim1.75\lambda$, which overestimates the value of $\Lambda\sim\lambda$ reported in \cite{Ohzono.2012} by almost a factor of $2$.

\section{Conclusions}
\label{sec:conclusions}

We investigated numerically the zigzag line defect that appears in sinusoidal channels filled with nematic liquid crystals, reported in \cite{Ohzono.2012}. 
Our study reveals that the line defect is rotated from its original position driven by a twist distortion that occurs in the region between the defect and the sinusoidal substrate when the twist elastic constant is smaller than the bend and splay elastic constants. This rotation leads to a decrease of the distance  between the defect and the surface of the channel until the alignment imposed by the substrate dominates. As a result the defect is rotated in the opposite direction in order to accommodate twist distortions. This process leads to a periodic zigzag configuration, where linear segments oriented by twist distortions of different handedness join, in the middle of the channel, at kinks. A detailed analysis of the nematic configuration at the kinks reveals that they are dominated by splay distortions. 

We also studied the stability of the zigzag defect for different substrate wavelengths and different elastic anisotropies $L_2/L_1$ (within a two-constant approximation). We found that the zigzag defect is stable when $L_2/L_1\gtrsim 1.5$ and $\lambda\gtrsim 0.8\mu$m. Outside of that region the defect adopts a straight line configuration. The elastic anisotropy threshold is higher than that predicted by the Frank-Oseen elastic theory by almost a factor of 2. We attribute this discrepancy not only to additional approximations of the analytical estimate, but also to the fact that it neglects the substrate releif.

Finally, we looked at the dependence of the zigzag periodicity $\Lambda(\lambda)$ on the wavelength of the substrate for $\lambda<5\mu$m. We found that for small values of $\lambda$ the period of the zigzag converges to a constant value $\Lambda \sim 1.65\mu$m. As $\lambda$ increases the defect increases (over the twist regions) and as a result $\Lambda$ increases. We found that $\Lambda$ increases smoothly and for $\lambda>4\mu$m depends linearly on the wavelength of the channel.

Throughout our study we fixed the ratio of the height $A$ and the wavelength $\lambda$ of the channels, $A/\lambda\simeq 0.1$, which is equivalent to considering substrates with the same roughness. The value chosen is similar to that used in the experiment of Ohzono and Fukuda \cite{Ohzono.2012}. We stress that increasing the roughness stabilizes the straight line configuration.

\section*{Acknowledgements}
We acknowledge the financial support of the Portuguese Foundation for Science and Technology (FCT) under the contracts numbers UID/FIS/00618/2013 (NMS and MMTG) and EXCEL/FIS-NAN/0083/2012.
JMRE also acknowledges partial financial support from Spanish Ministerio de Econom\'{\i}a y Competitividad through grant no. FIS2012-32455, and Junta de Andaluc\'{\i}a through grant no. P09-FQM-4938, all co-funded by the EU FEDER.

\section*{References}

\bibliography{references}

\end{document}